# Micro-Raman and resistance measurements of epitaxial La$_{0.7}$Sr$_{0.3}$MnO$_3$ films

V. A. Dediu, J. López, F. C. Matacotta, P. Nozar, G. Ruani, R. Zamboni, C. Taliani

Istitutto di Spetroscopia Molecolare, CNR, via Gobetti 101, 40129 Bologna, Italy

The Channel-Spark method was used for deposition of highly oriented ferromagnetic La$_{0.7}$Sr$_{0.3}$MnO$_3$ films on NdGaO$_3$ substrates. It was found that additional oxygen decreases the film quality suppressing the Curie temperature and metal-insulator transition below the room temperature. To achieve the best quality of the films the samples were either annealed in high vacuum at deposition temperature or even deposited in argon atmosphere with no oxygen annealing. For such films the resistive measurements showed a metallic behaviour in the interval 10-300 K in accordance with the high Curie point (T$_c \geq$ 350 K). Micro-Raman analysis indicate that the La$_{0.7}$Sr$_{0.3}$MnO$_3$ films are well ordered, while some outgrowths show stoichiometrical deviations.

The Channel-Spark method (CS) has been successfully used previously for the deposition of the films of the 123 superconducting compounds, providing a film quality similar to that obtained by Laser ablation [1-2]. We report in this article the use of CS for the deposition of the Colossal Magnetoresistance (CMR) manganite La$_{0.7}$Sr$_{0.3}$MnO$_3$.

The CS system consists of pulsed electron beam generator and the heater-sample-target system. The electron pulses are created by an electron avalanche process in the zero-gradient field region of the hollow cathode [1]. Because of the radial component of the electrical field in the hollow cathode, the electron avalanches are focused to the vertical axis and escape as a tiny beam to the high voltage (up to 20 kV) acceleration section. In spite of the carrier repulsion, the magnetic self-pinching provides high-current densities [3]. The resulting pulses have energy density of about 1 J cm$^{-2}$ and duration about 100 ns.

For the manganite deposition both oxygen and argon were used as working gases. Gas pressure was kept constant during process - P = 3 10$^{-2}$ mBar. Sample was heated by a stainless steel heater to temperatures about 700-750 °C. Stoichiometrical targets were produced by a conventional pellets sintering [4] at 1200 °C or 1500 °C, providing densities of about 5.5 g/cm$^3$ and 6 g/cm$^3$ respectively. Target - sample distance was about

40 mm. Acceleration voltage 10-15 kV provided at this pressure deposition rates ranging from 0.05 Å/pulse to 0.15 Å/pulse. Pulses repetition rate was kept at ~ 3 Hz. The film thicknesses (d=500-2500 Å) were measured by a conventional $\alpha$ - step technique. The XRD analysis indicate that the films are strongly c-axis oriented. Some very small peaks could be identified as a minority growth fraction along (hh0).

Our main goal was to investigate the role of oxygen during film growth as well as the necessity of postannealing process. Differently from superconducting cuprates the manganites do not need additional oxygenation. Moreover, oxygenation can create defects like cation vacancies due to a lattice instability to the incorporation of oxygen in La-Mn-O system [5]:

$$LaMnO_{3+\delta} -> La_{3/(3+\delta)}Mn_{3/(3+\delta)}O_3 \qquad (1)$$

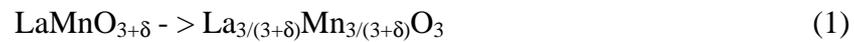

Three different procedure were used for the $La_{0.7}Sr_{0.3}MnO_3$ film deposition-annealing-cooling process. First procedure was quite similar to the cuprate deposition: after the deposition at 700 °C and oxygen pressure of 3 $10^{-2}$ mBar samples were cooled to 450 °C and the pressure was increased to 800 mBar for 30-60 min. In the second procedure the annealing was performed at deposition temperature in high vacuum, then the sample was quenched to room temperature. Third procedure involved no use of oxygen at all, the argon being used as a working gas (3 $10^{-2}$ mBar) at 700 °C with immediate quenching after deposition. The highest film quality was found for the second and third procedure. Fig. 1 shows the temperature dependence of the resistance for three 1000 Å films deposited by different procedures on NdGaO$_3$ substrates (film 1, film 2, film 3 respectively). We shall note that the Curie temperature in $La_{0.7}Sr_{0.3}MnO_3$ is ~350 K and nearly coincides with insulator-metal (MI) transition (maximum on R(T)). As can be seen, the films deposited without additional annealing in oxygen are metallic with the MI transition above 300 K. On the other hand the film annealed in oxygen shows a lower $T_C$ and $T_{MI}$ (~ 250 K), having also a region of increasing R at low temperature. Such regions are characteristic for the tunneling connection between ferromagnetic grains. The residual low temperature resistivity of this film is significantly higher than that of two others, the room temperature resistivity being quite similar for all films: $\rho \approx 10$ m$\Omega$ cm. These results can be explained supposing that the additional oxygen atoms create defected regions of La-Sr-Mn-O or La-Mn-O type with either lower ferromagnetic or even antiferromagnetic order. These material can be placed at the grain boundaries, separating the stoichiometric domains via tunneling. At low temperatures the tunneling is dumped leading to the observed resistance increasing. We can conclude that not only the oxygen annealing but even the use of oxygen as a working gas is not necessary for the deposition of high quality stoichiometrical films of $La_{0.7}Sr_{0.3}MnO_3$. This means that the target molecules are not destroyed by Channel Spark ablation, but move as separate molecules and small clusters of a stoichiometrical compound.

Micro-Raman studies were performed on both targets and thin films. The Raman scattering measurement were recorded in back scattering geometry by using a Renishaw RM 1000 system equipped with a DMLM series Leica microscope with a spectral resolution of approximately 2 $cm^{-1}$. An objective of x50 allowed us to perform Raman scattering, in confocal condition, on microcrystals smaller than 2 μm in diameter. The $Ar^+$ laser emission at 488 nm was used as excitation source. Measurements were performed by using an *z(xx)z̲* geometry. As the films are twinned and we cannot distinguish between *a* and *b*, this means that our spectra represent a sum of the two scattering matrix elements *xx* and *yy*.

The Raman spectra for the target (Figure 2) are in agreement with the measurements performed on single crystal [6], and indicate a superposition of three different components: the phonon peaks at ~200 $cm^{-1}$ and 425 $cm^{-1}$, a very broad peak (from low energy to ~700 $cm^{-1}$), and a strong enough background easily visible at high energies. The peak at 425 $cm^{-1}$ is intense and sharp, while the peak at 200 $cm^{-1}$ is strongly obscured by the experimental setup cut-off frequency, making difficult the estimation of its position and intensity.

In order to estimate the film quality we shall mainly analyze the phonon peaks. Differently from orthorhombic $LaMnO_3$ system [7], the phonon peaks of rombohedricaly distorted perovskite $La_{(1-x)}M_xMnO_3$ are not yet well identified. Nevertheless, by analogy the 200 $cm^{-1}$ and 425 $cm^{-1}$ modes should reflect the rare-earth/$MnO_6$ vibration mode and the apical oxygens bending mode of $MnO_6$ octahedra respectively.

Two different film's Raman spectra, for films 2 and 3, are presented in Figure 3 and Figure 4 respectively. We shall note that these spectra are typical for the films prepared by the second and third procedure. The films prepared by the first procedure showed no resolved Raman peaks, except the substrate ones. From preliminary AFM study results that all the films consist mainly of a flat background with a roughness below few unit cells, and some large outgrowths with dimensions, that can duplicate the film thickness. Such outgrowths are typical for the perovskite films, and were in details studied on 123 system thin films [7]. The spectra for both films consist of a sum of $NdGaO_3$ and $La_{0.7}Sr_{0.3}MnO_3$ spectra (the 425 cm$^{-1}$ energy is indicated by dotted line). The $La_{0.7}Sr_{0.3}MnO_3$ mode is well visible, although it is slightly obscured by the substrate 405 cm$^{-1}$ mode for the film 2. As in the film 3 both the smooth film part and the outgrowths show the presence of 425 cm$^{-1}$ mode, indicating similar composition, the film 2 shows for the outgrowths an additional mode placed at ~ 640 cm$^{-1}$. One can also see that the intensity of this broad peak is inversely proportional to that of the 425 cm-1 mode (the outgrowths 1 and 2). The 640 cm$^{-1}$ mode is close energetically to the stretching modes of the orthorhombic perovskites like $Pr_{1-x}Ca_xMnO_3$ or $LaMnO_{3+\delta}$. We suppose the origin of these orthorhombic-like phase distortion to be caused by additional oxygen incorporation in structure and cation vacancies formation. This indicates that the third procedure, where the deposition of the manganite is performed totally in argon with no any additional oxygen treatment, provides the best film quality and the highest homogeneity.

The wide band observed in bulk samples is present also in the films, but covers a wider spectral interval (up to 800 cm$^{-1}$). We suppose this band can be tentatively attributed to a multiphonon structure. Such behaviour can be observed around a local distortion induced in the crystal by the presence of a polaron as a consequence of the selection rules relaxation due to the break of the system periodicity. The increase of the band energy interval in the films can be caused by the substrate effect. Similar structures have been observed in IR reflectance measurements in light doped parent compounds of HT$_c$ perovskites [8]. A confirmation of such an interpretation arises also from our preliminary results on the temperature dependence of Raman spectra of the films (results will be published elsewhere).

In conclusion, three different procedures were used for the $La_{0.7}Sr_{0.3}MnO_3$ thin films deposition by Channel-Spark electron ablation. Resistive measurements and Micro-Raman investigation show that the best film quality is provided by the in-situ preparation in argon atmosphere. Additional oxygen treatments introduce the cation defects in the films, depressing the structural and magnetic order. We explain these results by the stoichiometrical cluster transfer of the material from target to the substrate.

We would like to acknowledge Martin Brinkman and Paolo Scardi for the AFM and XRD sample characterization. J. López thanks the Condensed Matter Section and the Training in Italian Laboratories Program of the Abdus Salam ICTP for the economic support, which allowed him to do this work.

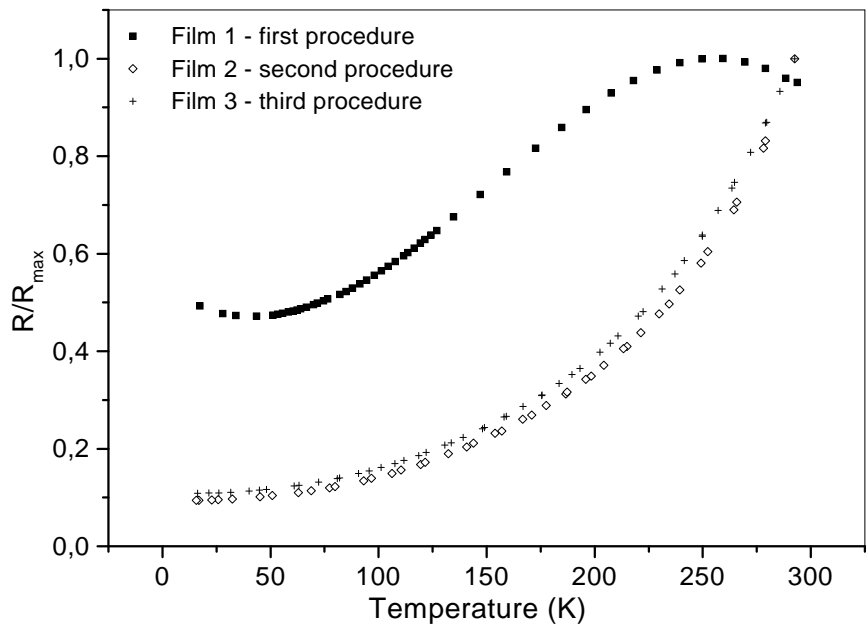

Fig. 1 The temperature dependencies of resistance for films 1, 2 and 3 normalized to maximum resistance value. Resistivity for all samples is about 10 mΩ cm at 300 K.

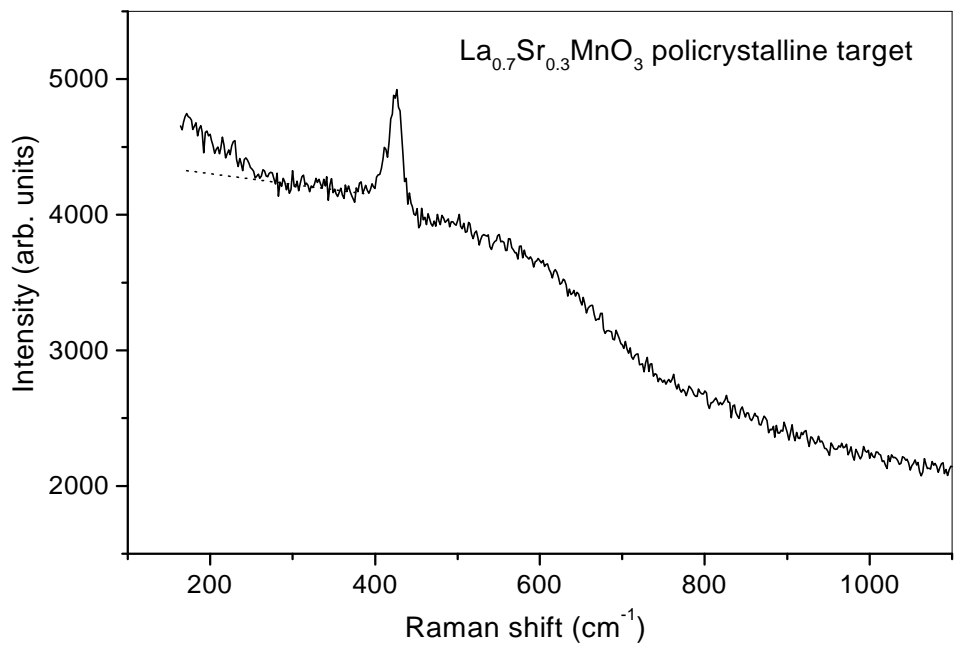

Fig. 2 Raman scattering spectrum for the $La_{0.7}Sr_{0.3}MnO_3$ target performed at room temperature. The short line is the guide for eyes and helps to separate the ~200 cm$^{-1}$ peak from the background.

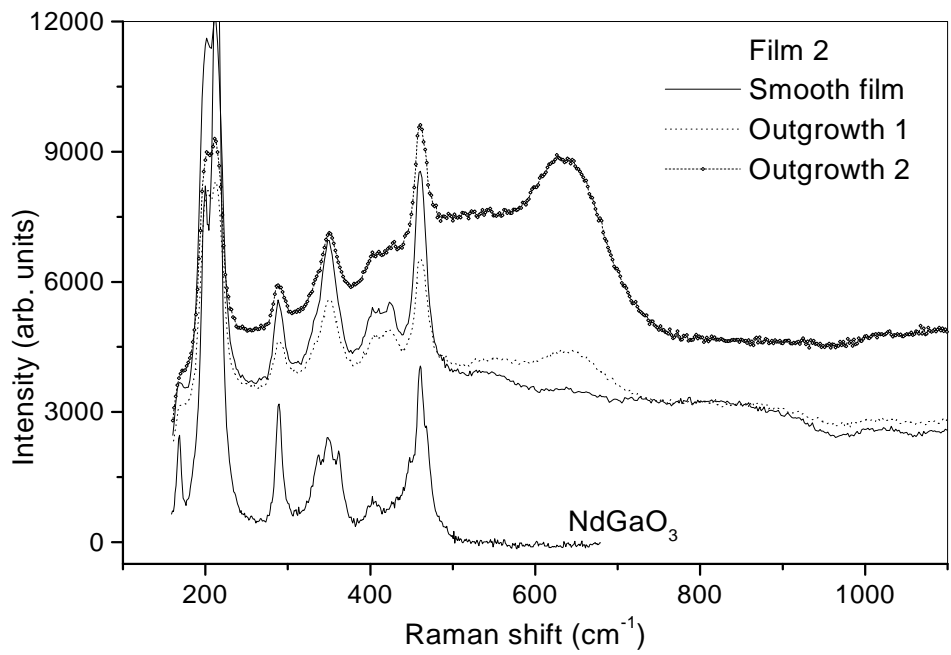

Fig. 3 Raman scattering spectrum for the film 2 and NdGaO3 substrate.

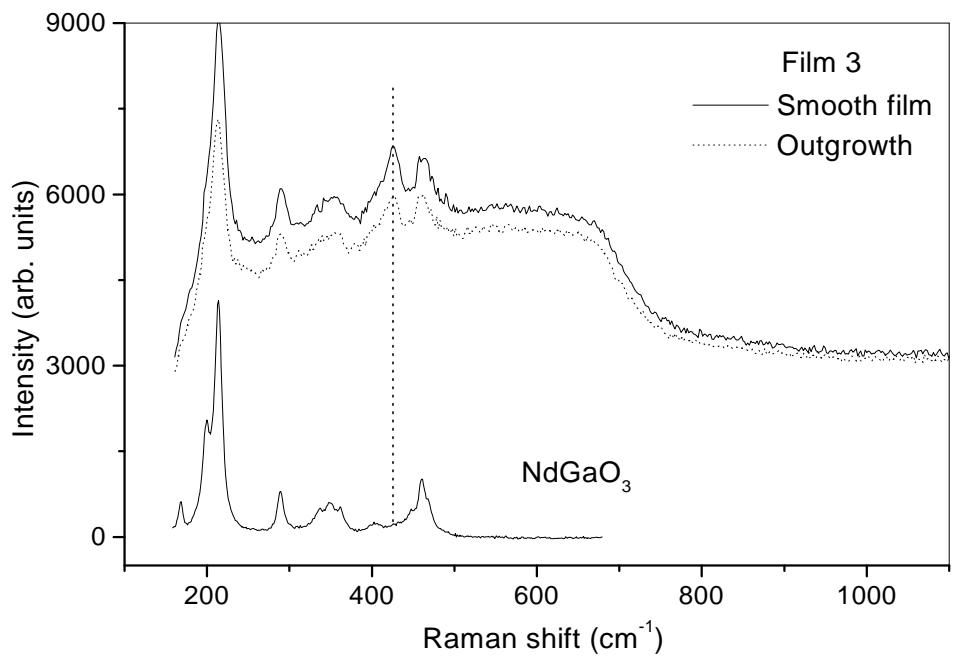

Fig. 4 Raman scattering spectrum for the film 3 and NdGaO3 substrate.